\documentclass[a4paper,11pt]{article}
\usepackage{pos}
\usepackage{slashed}
\usepackage{bbm}
\title{Confronting a Standard Model extension with a dark $U(1)$ gauge sector with the prediction for the W-boson mass}
 \ShortTitle{Confronting the Dark Abelian Sector Model with the prediction for the W-boson mass}

\author[a]{Stefan Dittmaier}
\author[a]{Jonas Rehberg}
\author*[a]{Heidi Rzehak}

\affiliation[a]{Albert-Ludwigs-Universit\"at Freiburg, %
        Physikalisches Institut, \\
        Hermann-Herder-Str. 3, 79104 Freiburg, %
        Germany}

\emailAdd{stefan.dittmaier@physik.uni-freiburg.de}
\emailAdd{jonas.rehberg@physik.uni-freiburg.de}
\emailAdd{heidi.rzehak@physik.uni-freiburg.de}

\abstract{The Dark Abelian Sector Model (DASM) is an extension of the Standard Model of particle physics with an additional spontaneously broken $U_\text{d}(1)$ gauge symmetry connected to a dark sector, i.e.\ the SM particles do not carry the corresponding charge.  In addition to the gauge boson resulting from the extra $U_\text{d}(1)$ gauge symmetry, the particle content is extended by a further Higgs boson, one Dirac fermion as well as right-handed neutrinos.  Employing the $U_Y(1)$ field-strength tensor as well as the SM Higgs mass operator (the only two singlet operators of the SM with dimension less than four) and the right-handed neutrino fields, we open three portals to the dark sector.  

After an introduction of the model, we discuss a renormalization scheme for the complete model with a special focus on the renormalization of the mixing angles. Finally, as an example of application, we present  the prediction for the W-boson mass derived from muon decay in the DASM.}

\FullConference{Loops and Legs in Quantum Field Theory (LL2024)\\
 14-19, April, 2024\\
Wittenberg, Germany\\}

\begin{document}
\maketitle

\section{Introduction}

While the results of collider experiments nicely conform with the predictions of the Standard Model (SM) so far, there are still open questions the SM cannot answer. One of these questions is what dark matter is made up of. In order to find hints for the actual realization of dark matter in terms of new particles, many different SM extensions are considered in the literature. One possibility that has been discussed is the extension of the SM gauge group by an additional $U(1)$~symmetry, see e.g.\ Refs.~\cite{Zp1, Hold, BabuZZpold, SchabWells, Wells, Tro, U1MW, Pel, Bento, U1DM}. In this proceedings contribution, we will present a particular $U(1)$-extended model with a dark sector. A more detailed discussion can be found in Ref.~\cite{Dittmaier:2023ovi}.

\section{The Dark Abelian Sector Model}

The Dark Abelian Sector Model (DASM) is an extension of the SM with an additional $U_\text{d}(1)$ gauge symmetry and a corresponding charge $\tilde{q}$ and gauge field $C^\mu$,
\begin{align}
 \mathcal L_{U_\text{d}} =  -\frac{1}{4}C^{\mu\nu}C_{\mu\nu}
\end{align}
with  $C_{\mu\nu}=\partial_\mu C_\nu-\partial_\nu C_\mu$. Kinetic mixing provides a portal to the SM in the gauge sector,
\begin{align}
  \mathcal L_{U_\text{d}, \mathrm{portal}} = - \frac{a}{2} B_{\mu \nu} C^{\mu \nu},
\end{align}
where $B_{\mu\nu}=\partial_\mu B_\nu-\partial_\nu B_\mu$ and $B^\mu$ the $U(1)$ hypercharge gauge field of the SM. 
The neutral gauge fields $W^{3, \,\mu}$, which is the third component of the $SU_L(2)$ gauge-field multiplet, $B^\mu$, $C^\mu$ mix to the mass eigenstates $A^\mu$, $Z^\mu$, and ${Z'}^{\mu}$ corresponding to the photon, the Z boson and a $Z'$ boson.

The mass of the $Z'$ boson is generated via spontaneous symmetry breaking. This requires an extension of the Higgs sector by a complex Higgs field $\rho$, which is a singlet under the SM gauge group and has a dark charge $\tilde{q}_\rho =1$, 
\begin{align}
\mathcal L_{\text{Higgs}} =\mu_2^2\Phi^\dagger\hspace{-2pt}\Phi + 2\mu_1^2\rho^\dagger\rho-\frac{\lambda_2}{4}(\Phi^\dagger\hspace{-2pt}\Phi)^2-4\lambda_1(\rho^\dagger\rho)^2-2\lambda_{12}\Phi^\dagger\hspace{-2pt}\Phi\rho^\dagger\rho, \label{eq:LHiggs}
\end{align}
where $\Phi$ is the SM-like Higgs doublet. The last term in Eq.~\eqref{eq:LHiggs} is the scalar portal which allows for interactions between the Higgs singlet and the SM fields. The neutral scalar components of the Higgs doublet and of the Higgs singlet can mix, and the resulting mass eigenstates are $h$, $H$.

In order to allow for neutrino masses, the DASM comprises right-handed neutrino fields for each generation, $\nu'^R_j$, $j =1,2,3$. Furthermore, a neutral Dirac fermion $f'_{\text{d}}$ with a dark charge $\tilde{q}_{\text{f}}=1$ is added. The corresponding fermion Lagrangian can be written as
\begin{align}
\mathcal{L}_\text{Fermion}={}&\mathcal{L}^\text{SM}_\text{Fermion}+\bar{f}'_{\text{d}}\left(\text{i}\slashed{D}_\text{d}-m_{\text{f}_{\text{d}}}\right)f'_{\text{d}}+\sum_{j=e,\mu,\tau}\left[\bar\nu'^{\text{R}}_{j}\text{i}\slashed{\partial}\nu'^{\text{R}}_{j}-\left(y_{\rho,j}\rho\bar{f}'^{\text{L}}_{\text{d}}\nu'^{\text{R}}_{j}+\text{h.c.}\right)\right]\nonumber\\
  &-\sum_{k,l=\text{e},\mu,\tau} \left(\bar{L}'^{\text{L}}_{k} G'^\nu_{kl}\nu'^{\text{R}}_{l} \Phi^\text{C}+ \text{h.c.}\right),
\label{eq:lag-ferm-exten}
\end{align}
where $L'_k$ is the left-handed lepton doublet, $G'^\nu$ the neutrino Yukawa matrix, and \mbox{$D_\text{d}^\mu =\partial^\mu+\text{i}\tilde{q}e_\text{d}C^\mu$} the covariant derivative of the dark gauge sector with the coupling constant $e_\text{d}$. The last term in the first line of Eq.~\eqref{eq:lag-ferm-exten} provides a portal between the dark sector and the SM with the Yukawa-type couplings $y_{\rho,j}$. In the following we assume that the masses of  SM-like neutrinos can be neglected. Taking this into account, we can simplify the mass matrix of the neutrinos, and it turns out that the physical effects can be described by only one of the SM-like neutrinos mixing with the dark fermion and, hence, by one Yukawa coupling $y_{\rho}$. This mixing  gives rise to two new neutrino-type mass eigenstates, one massless and one massive, in addition to two further massless neutrino mass eigenstates.

The parameters of the DASM comprise the SM ones
\begin{align}
g_1, g_2, g_3, \mu_2^2, \lambda_2, G^\ell, G^d, G^u, v_2,
\end{align}
where $g_1$, $g_2$, $g_3$ are the $U(1)$ hypercharge, the $SU(2)$ isospin, and the $SU(3)$ gauge coupling, $G^i$ with $i = \ell, d, u$ are the Yukawa coupling matrices of the charged leptons, the down-type and the up-type quarks, and $v_2$ denotes the vacuum expectation value of the Higgs doublet.
The additional parameters originating from the non-SM-like sector are summarized as
\begin{align}
 e_\text{d},  a,  \mu_1^2, \lambda_1, \lambda_{12}, y_\rho,  m_{\text{f}_{\text{d}}}, v_1,
\end{align}
with $v_1$ being the vacuum expectation value of the Higgs singlet.
These original parameters of the Lagrangian are replaced by the following set of input parameters for the SM parameters,
\begin{align}
\alpha_s, M_\text{W}, M_\text{Z}, \alpha_{\text{em}}, M_{H_{\text{SM}}}, m_f, V,
\end{align}
and for the parameters of the non-SM part,
\begin{align}
M_\text{Z'},
\gamma, M_{h'}, \alpha, \lambda_{12}, m_{\nu_4}, \theta_r, v_1,
\end{align}
with $\alpha_s$ and $\alpha_{\text{em}}$ being the strong and the electromagnetic coupling constants, respectively, $M_\text{W}$, $M_\text{Z}$, $M_\text{Z'}$ denoting the gauge-boson masses, $M_{H_{\text{SM}}}$, $M_{h'}$ the SM-like and the non-SM-like Higgs-boson mass, $m_f$ the fermion masses, $m_{\nu_4}$ the mass of the fourth neutrino, and $\gamma$, $\alpha$, $\theta_r$ the gauge-boson, the Higgs-boson, and the neutrino mixing angles, respectively, and $V$ the CKM matrix.
These input parameters are also those that enter our renormalization procedure where we define
\begin{itemize}
\item[$\bullet$] masses and fields via on-shell conditions,
\item[$\bullet$] mixing angles via ratios of on-shell form factors or as $\overline{\text{MS}}$ parameters, 
\item[$\bullet$] the electric charge as coupling of fermion--photon interaction in the  Thomson limit, 
\item[$\bullet$] $\lambda_{12}$ as $\overline{\text{MS}}$ parameter.
\end{itemize}
The CKM matrix has been chosen as unity matrix throughout our work. The treatment of the vacuum expectation values included above are connected to the treatment of the tadpole parameters for which we offer several alternatives including the gauge-invariant vacuum expectation value scheme (GIVS), see Refs.~\cite{SDTad, Dittmaier:2023ovi}.

Since the gauge-boson mixing angle $\gamma$ enters the calculation of the W-boson mass, we sketch our renormalization condition for $\gamma$ here: Following the idea of Ref.~\cite{SDMixingAngles}, we introduce a 
 ``fake fermion'' field~$\omega_\text{d}$,
\begin{align}
{{\mathcal L_{\omega_{\text{d}}} ={{ \bar{\omega}_\text{d}}}
    \Bigl[
        \text{i} \slashed{\partial} - \tilde{e} {{\tilde{q}_\omega}}(s_\gamma  \slashed{Z} + c_\gamma \, \slashed{Z'})
      - m_{\omega_\text{d}}\Bigr]{{\omega_\text{d}}}}},
\end{align}
where we introduced the short-hand notation $s_\gamma = \sin \gamma$, $c_\gamma = \cos \gamma$ and  $\tilde{e} = e_\text{d}(1 - a^2)^{-\frac{1}{2}}$. The vertices between the fake fermions and the gauge bosons $Z$, $Z'$ vanish for vanishing charge of the fake fermion, $\tilde{q}_\omega$,
\begin{equation}
\setlength{\unitlength}{1pt}
\raisebox{-18pt}{\includegraphics{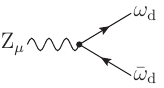}} 
= -\text{i}{{\tilde{q}_{\omega}}}\tilde{e}s_\gamma\gamma_\mu ,\hspace{0.8cm}
  \raisebox{-18pt}{\includegraphics{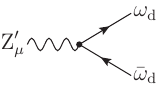}} 
= -\text{i}{{\tilde{q}_{\omega}}}\tilde{e}c_\gamma\gamma_\mu.\nonumber
\end{equation}
The amplitudes corresponding to the decay of the gauge bosons $Z$, $Z'$ can be expressed in terms of the form factors
 $\mathcal{F}^{Z\bar{\omega}_\text{d}{\omega_\text{d}}}$ and $\mathcal{F}^{Z'\bar{\omega}_\text{d}{\omega_\text{d}}}$ as
\begin{align}
  \mathcal{M}^{\text{Z}\rightarrow\bar{\omega}_\text{d}{\omega_\text{d}}}= [\bar{u}_{\omega_\text{d}}\slashed{\varepsilon}v_{\omega_\text{d}}]_{Z} {{\mathcal{F}^{Z\bar{\omega}_\text{d}{\omega_\text{d}}}}},\qquad\mathcal{M}^{\text{Z}'\rightarrow\bar{\omega}_\text{d}{\omega_\text{d}}}= [\bar{u}_{\omega_\text{d}}\slashed{\varepsilon}v_{\omega_\text{d}}]_{Z'} {{ \mathcal{F}^{Z'\bar{\omega}_\text{d}{\omega_\text{d}}}}}.
\end{align}
For the renormalization condition defining $\gamma$, the ratio of the form factors is fixed in such a way that loop contributions vanish,
\begin{align}
{{
\lim_{{\tilde{q}_{\omega}\rightarrow 0}}\frac{{{\mathcal{F}^{Z\bar{\omega}_\text{d}{\omega_\text{d}}}}}}{{{\mathcal{F}^{Z'\bar{\omega}_\text{d}{\omega_\text{d}}}}}}\stackrel{ !}{=}\frac{{{\mathcal{F}_\text{LO}^{Z\bar{\omega}_\text{d}{\omega_\text{d}}}}}}{{{\mathcal{F}_\text{LO}^{Z'\bar{\omega}_\text{d}{\omega_\text{d}}}}}} = \frac{s_\gamma}{c_\gamma}
}}.
\label{eq:gamma-rc}
\end{align}
The ratio of the form factors at NLO can be expressed by
\begin{align}
  \frac{{{\mathcal{F}_\text{NLO}^{Z\bar{\omega}_\text{d}{\omega_\text{d}}}}}}{{{\mathcal{F}_\text{NLO}^{Z'\bar{\omega}_\text{d}{\omega_\text{d}}}}}}=
      {}\frac{{{\mathcal{F}_\text{LO}^{Z\bar{\omega}_\text{d}{\omega_\text{d}}}}}}{{{\mathcal{F}_\text{LO}^{Z' \bar{\omega}_\text{d}{\omega_\text{d}}}}}}\biggl[&1+\frac{\delta s_\gamma}{s_\gamma}-\frac{\delta c_\gamma}{c_\gamma}
+\frac{1}{2}\left( \delta Z_{ZZ}- \delta Z_{Z'Z'}+\frac{c_\gamma}{s_\gamma} \delta Z_{Z'Z}-\frac{s_\gamma}{c_\gamma} \delta Z_{ZZ'}\right)\nonumber\\
&+\delta^{Z\bar{\omega}_\text{d}{\omega_\text{d}}}_{\text{loop}}-\delta^{Z'\bar{\omega}_\text{d}{\omega_\text{d}}}_{\text{loop}}\biggr], \label{eq:NLOformfac}
\end{align}
where the field renormalization constants are introduced via the renormalization transformations of the fields
\begin{align}
  \begin{pmatrix}A_{0}\\Z_{0}\\Z'_0\\\end{pmatrix}&={}\left(\mathbbm{1}_3 +\frac{1}{2}\delta\mathrm{Z}_\text{V}\right)\begin{pmatrix}A\\Z\\Z'\\\end{pmatrix}\qquad \text{with} \qquad \delta\mathrm{Z}_\text{V}=\begin{pmatrix} \delta Z_{AA}&\delta Z_{AZ}&\delta Z_{AZ'}\\\delta Z_{ZA}&\delta Z_{ZZ}&\delta Z_{ZZ'}\\\delta Z_{Z'A}&\delta Z_{Z'Z}&\delta Z_{Z'Z'}\\\end{pmatrix}.
\end{align}      
The renormalization constant $\delta \gamma$ corresponding to the gauge-boson mixing angle is hidden in $\delta s_\gamma = c_\gamma \delta \gamma$ and  $\delta c_\gamma = -s_\gamma \delta \gamma$. The relative vertex contributions 
$\delta^{Z\bar{\omega}_\text{d}{\omega_\text{d}}}_{\text{loop}}$, $\delta^{Z'\bar{\omega}_\text{d}{\omega_\text{d}}}_{\text{loop}}$ vanish in the limit of vanishing $\tilde{q}_\omega$. Substituting Eq.~\eqref{eq:NLOformfac} into Eq.~\eqref{eq:gamma-rc} and solving for $\delta \gamma$ results in
\begin{align}
  {{\delta\gamma
      =\frac{1}{2}s_\gamma c_\gamma\left(\delta Z_{Z'Z'}-\delta Z_{ZZ}\right)+\frac{1}{2}\left(s_\gamma^2\delta Z_{ZZ'}-c_\gamma^2\delta Z_{Z'Z}\right)}},
  \end{align}
which only depends on field renormalization constants.

\section{The muon decay and the prediction of the mass of the W boson}

At leading order, the decay width of the muon in the DASM is given as
\begin{align}
  \Gamma^W_{\text{DASM}} = \frac{{{\alpha_{\text{ew}}^2}} m_\mu^5}{384
    \pi {{M_\text{W}^4 s_\text{w}^4}}}\biggl(1 - \frac{8 m_e^2}{m_\mu^2}\biggr)
\end{align}
with the sine squared of the weak mixing angle $s_\text{w}^2 = 1- M_\text{W}^2/(c_\gamma^2 M_\text{Z}^2 + s_\gamma^2 M_\text{Z'}^2)
$. Since the W~boson is very heavy in comparison to the muon and the electron mass, the muon decay can be well described within the Fermi theory, 
\begin{align}
  \Gamma^W_{\text{Fermi}}=  \frac{{G_F^2} m_\mu^5}{192 
    {\pi^3}}\biggl(1 - \frac{8 m_e^2}{m_\mu^2}\biggr),
\end{align}
with the Fermi constant $G_F$, which has been measured very precisely. Equating the two decay widths, the W-boson mass can be calculated in terms of the Fermi constant. Higher-order corrections change this relation to
\begin{align}
  \frac{{G_F}}{ \pi} = \frac{{{\alpha_{\text{ew}}}}}{{{\sqrt{2}}} {{M_\text{W}^2 s_\text{w}^2}}}(1 + \Delta r), 
\end{align}
where
\begin{align}
  \Delta r 
  = 2{{\delta Z_e}}-\frac{{{\delta s_\text{w}^2}}}{s_\text{w}^2}-\frac{{{\delta M_\text{W}^2}}}{M_\text{W}^2}+{{\frac{\Sigma_\text{T}^{WW}(0)}{M_\text{W}^2}}}+{{\delta_\text{vertex}}}+{{\delta_\text{box}^\text{massive}}}+{{\delta_\text{box}^{\gamma\text{W}}}}
\end{align}
with the first three terms comprising radiative corrections due to renormalization constants for the electric charge, the mass of the W boson, and the sine square of the weak mixing angle. The latter renormalization constant can be calculated from the renormalization constants of the gauge-boson masses and of the gauge-boson mixing angle. The fourth term accounts for propagator corrections with $\Sigma^{WW}_\text{T}$ being the transverse W self-energy. The last term $\delta_\text{box}^{\gamma\text{W}}$ denotes QED box contributions which can be taken over from the SM~\cite{hollikmuon}, since the QED part in the DASM and the SM are the same, while ${{\delta_\text{vertex}}}$ and ${{\delta_\text{box}^\text{massive}}}$ include vertex and box contributions in addition to the SM ones.

In order to eliminate the dependence on light-quark masses, we replace electromagnetic coupling constant $\alpha_{\text{em}}$ by $\alpha_\text{em} (M_\text{Z}^2)= \alpha_\text{em}(0)(1-\Delta \alpha_\text{em})^{-1} 
$. In addition, we resum terms proportional to top quark masses $\Delta \rho$ up to $\mathcal O(s_\gamma^2)$ and replace $ s_\text{w}^2$  by $\bar{s}_\text{w}^2=s^2_\text{w}+c_\text{w}^2\Delta \rho$. Splitting $\Delta r$ in \mbox{$\Delta r= \Delta \alpha_\text{em}-\frac{c_\text{w}^2}{s_\text{w}^2}\Delta \rho+\Delta r_\text{rem}$}~\cite{rhoparam, rhoparam2, deltaRcor2, deltaRcor3, deltaRcor5}, we obtain
\begin{align}
    G_\text{F}=\frac{\alpha_\text{em}(M_\text{Z}^2) \pi}{\sqrt{2}\bar{s}_\text{w}^2 M_\text{W}^2}\left(1+\Delta r_\text{rem}\right),
\end{align}
which can be used for calculating the W-boson mass at NLO. For the numerical evaluation, we make use of the best prediction in the SM \cite{MWmasseSM} by defining our best prediction for the W-boson mass within the DASM as
$M_\text{W}^\text{DASM}= M_\text{W}^\text{SM} + \Delta M_\text{W}$ where
$\Delta M_\text{W}=M_\text{W,NLO}^\text{DASM}-M_\text{W,NLO}^\text{SM}$ is the difference of the W-boson mass in the DASM and the SM at NLO.

\begin{figure}
\centering
                \includegraphics[width=16cm]{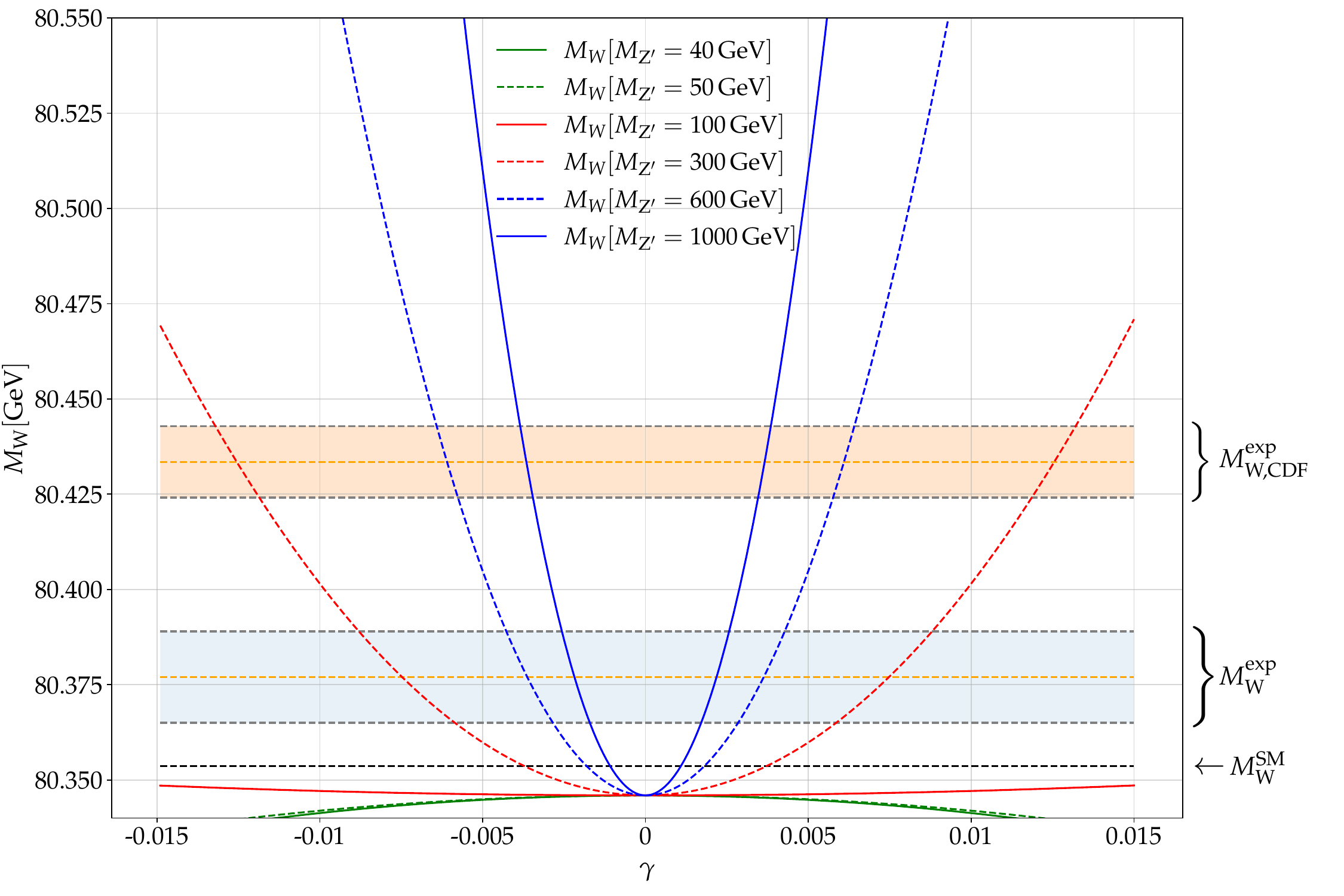}
                \caption{Predictions for $M_\text{W}$ in the DASM in dependence of $\gamma$ for various combinations of $M_{\text{Z}'}$. The best SM prediction is given by $M^\text{SM}_\text{W}=80.3536$ GeV \cite{MWmasseSM}, and the measured world average is \mbox{$M^\text{exp}_\text{W}=80.377\pm0.012$ GeV \cite{PDG}.} In addition, we show the result of the CDF experiment $M_\text{W,CDF}^\text{exp}= 80.4335 \pm 0.0094$ GeV \cite{CDFMW}, which has not been taken into account in the world average value quoted above. The figure is taken over from Ref.~\cite{Dittmaier:2023ovi}.
\label{fig:M_W_NLO}}
  \end{figure}
In Fig.~\ref{fig:M_W_NLO}, the W-boson mass is shown in dependence of the mixing angle $\gamma$ for different values of the mass of the $\text{Z}'$ boson in comparison to the SM prediction and the measured world average as well as the result of the CDF measurement. Clearly, depending on the parameter values, the DASM can predict a W-boson mass value that is closer to the measured one. However, in order to assess whether the DASM leads to a better performance than the SM, further observables have to be tested simultaneously.

\section{Conclusion and Outlook}

In this proceedings contribution, we have briefly reviewed the DASM~\cite{Dittmaier:2023ovi}, a model with three portals to a dark sector. A renormalization scheme for this model has been given and applied in the calculation of the W-boson mass. In order to evaluate whether the DASM is describing Nature better than the SM, we plan to fit several electroweak precision observables simultaneously and to investigate the most favourable region of parameter space.

\section*{Acknowledgements}

  H.R. thanks the organizers for the invitation to give this talk and for the interesting and enjoyable conference in Lutherstadt Wittenberg. Her research is funded by  the
Deutsche Forschungsgemeinschaft (DFG, German Research
Foundation)---project no.\ 442089526; 442089660.

\end{document}